# Detecting Deficiencies: An Optimal Group Testing Algorithm

Group testing has been the subject of continuing investigation since 1943, when the US army was looking for an economical way to screen recruits for syphilis by pooling minute blood samples [**1**]. Consider this scenario, commonly used for comparing results: We are testing samples of blood from one million people, attempting to locate the carriers of a disease known to be present in .01% of the population. With no further information, we assume each person has the same probability of testing positive (being diseased.) Instead of performing one million individual tests, we might economize by pooling samples. For example, if we test a mixture of one hundred samples, the result is likely to be negative, and ninety-nine tests will have been avoided. On the other hand, testing a mixture of all one million blood samples would be pointless, since several of the included samples would no doubt be positive and the test would yield no new information. If a requirement is to find all positive samples, can one specify an optimum procedure to do so, that is, one which always accomplishes this with the minimum expected number of tests? This requires a set of operating rules that specify not only how many samples should be grouped together in the initial tests, but also how to proceed within a group if it tests positive.

   Since its introduction in the above form, mathematically equivalent situations having to do with production lines, computer wiring, DNA screening and other areas, have produced a variety of approaches [**2**, pp. 1-5, **5,7**]. While blood testing is perhaps the version most easily grasped, we can state our assumptions more generically:

(i)    Tests are being administered to a large population of $n$ samples in order to determine exactly which samples test positive.

(ii)   Each sample has the same probability $p$ of testing positive.

(iii)  A test of a group of samples will register positive whenever one or more of the samples included in it are positive.

(iv)   The cost of testing a group is always one unit, regardless of group size.

(v)    If by previous tests a group is known to contain a positive sample, the group need not be tested, although subgroups within it might be.

(vi)   Once a group is tested, a subsequent test which includes samples within that

group cannot include samples outside that group. This is typically referred to as "nesting."

Within the constraints of these six assumptions there are several ways to define a valid search procedure. In this paper a valid search procedure is the following:

DEFINITION 1. A valid search procedure is defined as one that proceeds according to a predetermined sequence of tests, the only exception being the omission of any test known to be positive as a result of earlier tests in the sequence.

## Some points to be noted

Considering all the attention given to this now classic problem, one might well ask why a further approach is needed. Indeed, four high school calculus students produced four reasonably efficient methods for treating an example just slightly different from the one specified above [8]. The attraction of course lies not in trying to produce a new method to lower the current minimum number of tests in this example. Indeed, for any economic or industrial application we are probably close enough to optimum. The interest here is purely mathematical—the challenge of either producing an optimum procedure, or proving that one does not exist.

(Spoiler alert: before we reveal anything in the next paragraph, the reader might like to write down his/her idea for an optimum method—at least how large the initial groups to be tested should be.)

In contrast to previous top-down methods, this paper takes a constructive, algebraic approach whose initial investigation can be found in [9]. While the algorithm in this paper appears to be the first proven optimum procedure, it is restricted to the above six assumptions and our version of the valid search procedure. Although this is the most common formulation of the problem, there are others that we will discuss below. After presenting the algorithm (whose proof is detailed in the last section) we will explore the Fibonacci pattern that consistently emerges.

The traditional terminology for this problem uses "group" to signify a set of samples, rather than its standard algebraic definition.

Most papers in the literature consider a version of the problem in which a specific number of positive samples is already known to exist, rather than assuming the same probability for each sample. This is sometimes realistic, but more often not. In this paper we know the probability of any individual sample's registering positive, an assumption which carries less information than knowledge of the specific number of positives.

## Terminology and method in simple cases

**Two samples**  Consider the simplest non-trivial case, that of two samples. There are two possible procedures for testing. The first is to test each sample separately, which we represent by $xx$. The second is to test them together and then, if a defect is registered, test them separately. A common way of representing this would be a rooted tree graph with two branches. While the testing procedure for any number of samples can always be represented by a tree graph, this becomes cumbersome as that number increases. For this reason we will denote the second procedure—testing two samples together and then, if a defect is registered, testing them separately—by $\overline{xx}$, where the horizontal line signifies a test of the two samples together. When extended to situations with several samples, this representation will make the testing order readily decipherable. The horizontal lines can be seen as space-saving versions of the graph edges. To be precise:

DEFINITION 2. A test of several samples together, followed by individual tests of each sample, conventionally represented by the tree graph

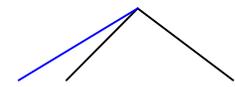

$x \quad x \quad . \quad . \quad . \quad . \quad . \quad x$ , will be represented by $\overline{xx.....x}$.  Similar representation will be used for all testing situations.

Since the expected number of tests for $xx$ is 2, regardless of p, we wish to know when $\overline{xx}$ is lower than 2. The four possibilities for a pair of samples are: [positive, positive], [positive, negative], [negative, positive] and [negative, negative]. For convenience, instead of the words "positive" and "negative" we can use the probabilities $p$ and $q$ themselves, where $q = 1 - p$. We will also place to the right of each possibility not the number of tests entailed by that possibility, but rather its excess above, or advantage below, 2 tests. Thus, for $\overline{xx}$:

$[p,p] \ 1 \quad [p,q] \ 1 \quad [q,p] \ 0 \quad [q,q] -1$

For example, [*p,q*] will require three tests—both together, then the first sample, then the second—so its excess is 1. [*q,p*] will require only two tests—both together, then the first sample, revealing that the second sample is *p*.

We now compute $v[\overline{xx}]$, the "value" of the procedure $\overline{xx}$, that is, the expected number of tests minus 2. (The general definition of value is given in section 3.) Multiplying the probability of each of the four possibilities above by its excess or advantage, we compute:

$$v[\overline{xx}] \;=\; p^2 + pq + 0 - q^2 \;=\; (1-q)^2 + q(1-q) - q^2 \;=\; 1 - q - q^2$$

If the value is negative, then $\overline{xx}$ is the better choice. If the value is positive, then $xx$ is better. Setting the above polynomial in *q* equal to zero, and confining ourselves to the interval [0,1], we find that $\dfrac{\sqrt{5}-1}{2}$ or $\phi$, the reciprocal of the golden mean, is the dividing point. If $q > \phi$ ($\approx .618$), then $\overline{xx}$ is better than $xx$, that is, has a lower value.

Since group testing is applicable only if *p* is quite small (*q* near 1) we can assume from now on that $q > \phi$. In addition, it will be convenient to write all of our equations, inequalities and expressions in terms of *q*.

**Three samples**  Testing three samples individually is represented by $xxx$ and, by analogy with the case of two, is given the value 0. Testing two together and one separately is represented by either $x\overline{xx}$ or $\overline{xx}x$, and from the results of section 3.1 we see that these have the same value, $v[\,x\overline{xx}\,] = v[\,\overline{xx}x\,] = 1 - q - q^2$. (Here the value is the expected number of tests minus 3, since testing separately would require three tests.)

One might perform $\overline{xxx}$, that is, test all three together and then, if necessary, test individually. Omitting the computation, $v[\overline{xxx}] = 1 - q^2 - 2q^3$.   A comparison of $v[\overline{xxx}]$ and $v[\,\overline{\overline{xx}x}\,]$ shows that $\overline{xxx}$ can never be optimal.

## *n* samples

 DEFINITION 3.  For any number of samples, *n*, a structure is defined as an arrangement of *x*'s and horizontal lines, where each line represents a collective test of all the samples beneath it. Tests are conducted  from top to bottom and left to right.

A structure unambiguously describes the procedure for testing *n* samples. For instance, an inclusive test of four samples, followed by a test of the first alone, then the last three, then the second alone, then the last two, then the third and the fourth—each test performed only when its outcome is not already known from previous tests—would be represented by $\overline{\overline{xx}xx}$. (There will be no ambiguity in context if we use the word "test" to mean either an actual test, or the line signifying a test.)

DEFINITION 4. The value of a structure, v[ ], is defined as its expected number of tests minus n, the number of samples it includes.

Our goal in the next three sections, stated once more, is to produce an algorithm which, for any given *n* and *q*, generates the structure with the lowest expected number of tests. As with the cases of *n* = 2 and *n* = 3, this is equivalent to finding the structure of least value.

DEFINITION 5. Test A is said to include, or is inclusive of, test B if all samples included in test B are included in test A. An all-inclusive test on a structure is defined as a test which includes all of its samples, and is represented by a line above all other lines in the structure.

What change in value occurs when an all-inclusive test is added to a given structure that lacks one? That is, what is v[the structure with the all-inclusive test] —v[the structure without the all-inclusive test]? Here we assume that if the all-inclusive test registers positive we continue testing exactly as we would have tested without it.

Consider structure #1, represented by $\overline{n_1}\,\overline{n_2}......\overline{n_k}$ , $\sum_{i=1}^{k} n_i = n$. In this structure there are *k* groups of samples, and we are representing each group simply by the number of samples in that group, even though each group may have its own substructure. We need not specify these substructures, for we will soon see that the value of adding an all-inclusive test above all *k* groups depends only upon the number of groups, *k*, and the number of samples, $n_i$ , in each group, not upon their precise substructures. Thus $\overline{n_i}$ represents a test of $n_i$ *x*'s with an unspecified substructure. If we add an all-inclusive test to structure #1 we will produce structure #2, $\overline{\overline{n_1}\,\overline{n_2}.....\overline{n_k}}$ .

Since we have added a single all-inclusive test to structure #1 to obtain structure #2, the value appears to increase by 1. However, should all the samples in the first $(k–1)$ groups be negative and some sample in group $k$ be positive, there would be no need to test group $k$ since, after testing the (k-1) groups which are negative, group $k$ would have to be positive. In this case the value would not increase by one, but remain the same. Thus, from an increase of 1 we should subtract the probability that this might occur, namely $(q^{\sum_{i=1}^{k-1} n_i})(1 - q^{n_k})$ or $(q^{n-n_k})(1 - q^{n_k})$. In addition, should all $n$ samples be negative, we would not have to perform any of the tests included in the $k$ groups. This would occur with probability $q^n$, and therefore we must subtract a further $kq^n$. The total additional value as a result of adding an all-inclusive test is then

$$1 - (q^{n-n_k})(1 - q^{n_k}) - kq^n \ = \ 1 - q^{n-n_k} - (k-1)q^n. \tag{1}$$

For convenience we refer to this last expression, $1 - q^{n-n_k} - (k-1)q^n$, as $E_1$.

What we have just shown is:

THEOREM 1. *The value of adding an all-inclusive test above a structure consisting of k groups depends only upon the number of groups, k, and the number of samples, $n_i$, in each group; it is independent of the precise substructures of these k groups.* ∎

We now have a rapid method for computing the value of any structure. Instead of going through the $2^n$ possibilities, as we did for $n = 2$ and 3, we simply begin with $n$ individual samples (value equal zero) and consider the tests one by one as they grow more inclusive. With each test we add the amount given by $E_1$, an amount we can call the value of the test. When no more tests remain we have calculated the total value of the structure.

Note that the order in which we calculate the value of a structure is the reverse of the order in which we execute the physical tests represented by that structure.

## Two theorems concerning optimal structures

DEFINITION 6. If a test has a negative value it is said to be advantageous; if it has a positive value it is said to be disadvantageous.

The two theorems in this section will prove that, in building a structure on *n* samples from the least inclusive tests up through the more inclusive—with the intention of finding the structure of least value—a disadvantageous test should never be added, and a test should never be applied to more than two tests immediately below it. That is, the optimal structure will contain only advantageous tests, and the structure will be binary, equivalent to a binary rooted tree.

THEOREM 2. *In building any structure, optimal or not, if a test is immediately disadvantageous, it can never be eventually advantageous.*

This theorem states that if a test has a positive value it should not be added. It is tempting to speculate that a disadvantageous test might eventually prove advantageous once inclusive tests are added above it, but this is never the case.

THEOREM 3. *Within an optimal structure, any test on m samples (m > 1) must include precisely two tests directly beneath it, where these two tests together include those same m samples. Thus the structure* $\overline{\overline{m_1 m_2 ..... m_k}}$ *could appear in an optimal structure only if k = 2.*

It should be noted that only now, using Theorems 2 and 3, is it possible to formally prove that for any *n*, if $q \leq \phi$, the structure without any group test at all is optimal.

## The optimal search algorithm

We can now specify a recursive algorithm to obtain the optimal structure on *n* samples. The structure thus obtained determines, in nesting fashion, the physical tests to be performed in the actual search. The proof that it is optimal is presented in the final section.

DEFINITION 7.
   a) $O_q(r)$ is defined as the optimal structure on *r* samples, each with the same probability *q* of being negative; $V_q(r)$ is its value.
   b) The union of two structures, $S_1 \bigcup S_2$, is defined as the structure obtained by placing $S_2$

to the right of $S_1$ and considering this a new structure.

c) $\overline{S_1 \bigcup S_2}$ is defined as $S_1 \bigcup S_2$ with the addition of an inclusive test of all samples.

d) $A_1$ is defined as the set of structures defined by $\{ O_q(n') \bigcup O_q(n-n') : 1 \le n' \le \lfloor \frac{n}{2} \rfloor \}$

e) $A_2$ is defined as the set of structures defined by $\{ \overline{O_q(n') \bigcup O_q(n-n')} : 1 \le n' \le \lfloor \frac{n}{2} \rfloor . \}$

THEOREM 4. *$O_q(r)$ is the structure of least value $V_q(r)$, selected from all structures included in $A_1$ or $A_2$ .*

Note that the set $A_2$ consists of the same $\lfloor n/2 \rfloor$ structures as those in $A_1$, with an additional all-inclusive test. Thus, utilizing the expression $E_1$ above, if the value of the $n'$-th structure in $A_1$ is $V_q(n') + V_q(n-n')$, the value of the $n'$-th structure in $A_2$ is $V_q(n') + V_q(n-n') + 1 - q^{n'} - q^n$ .

To briefly illustrate, imagine that in our recursive search for $O_q(7)$ we already know the optimal structures $O_q(n)$ and their values $V_q(n)$ for n $\le$ 6. We now choose the minimum of the values of the six structures $O_q(1) \bigcup O_q(6)$ ; $O_q(2) \bigcup O_q(5)$ ; $O_q(3) \bigcup O_q(4)$; $\overline{O_q(1) \bigcup O_q(6)}$ ; $\overline{O_q(2) \bigcup O_q(5)}$ ; $\overline{O_q(3) \bigcup O_q(4)}$ . This is $V_q(7)$, and the structure that produced it is $O_q(7)$.

In summary, we begin by obtaining $O_q(2)$ and its value $V_q(2)$. From this we obtain $O_q(3)$ and its value $V_q(3)$, and so on recursively until we reach $O_q(n)$ and its value $V_q(n)$. A structure of tests is thus recursively built, always combining exactly two tests, until a further test is no longer advantageous.

In practice, computing time can be conserved by noting that since a test can never include more than two tested groups immediately beneath it, its value over $n$ samples will be of the form $1 - q^m - q^n$ . This has a minimum of $1 - q - q^n$ , when $m = 1$, so we never need consider a test of more than $n_{max}$ samples, where $n_{max}$ is the greatest $n$ such that $1 - q - q^n \le 0$. Solving $1 - q - q^u = 0$, $u = \log(1-q)/\log(q)$, so that $n_{max} = \lfloor \log(1-q)/\log(q) \rfloor$.

# Computational results and the conjectured Fibonacci pattern

**A typical example**   In an optimal structure, if the size of the population, $n$, is sufficiently great, the number of samples included the most inclusive test is a function of $q$ only; $n$ is irrelevant. For example, in an optimal structure for $q = .9999$, the most inclusive tests are in groups of 6765. For a population with $n > 6765$, a test which included more than 6765 samples would always be disadvantageous. To continue this particular example, if any such group of 6765 samples registers positive, the subgroups to be tested should have sizes 2584 and 4181. Successive divisions, if positives are registered, are listed in Table 1, together with the expected number of tests. With $q = .9999$ and $n = 1,000,000$ the program took several minutes to run using Maple 10 on Macintosh, OS X.

TABLE 1: Successive divisions and expected number of tests

| size $n$ | expected #tests | division | size $n$ | expected #tests | division | size $n$ | expected #tests | division |
|---|---|---|---|---|---|---|---|---|
| 3, | 1.000699960, | 1, 2 | 26, | 1.017892624, | 8, 18 | 49, | 1.040072616, | 15, 34 |
| 4, | 1.001199900, | 1, 3 | 27, | 1.018792024, | 8, 19 | 50, | 1.041071609, | 16, 34 |
| 5, | 1.001699840, | 2, 3 | 28, | 1.019691384, | 8, 20 | 51, | 1.042070590, | 17, 34 |
| 6, | 1.002299730, | 2, 4 | 29, | 1.020590715, | 8, 21 | 52, | 1.043069521, | 18, 34 |
| 7, | 1.002899610, | 2, 5 | 30, | 1.021490155, | 9, 21 | 53, | 1.044068482, | 19, 34 |
| 8, | 1.003499490, | 3, 5 | 31, | 1.022389556, | 10, 21 | 54, | 1.045067394, | 20, 34 |
| 9, | 1.004199300, | 3, 6 | 32, | 1.023288926, | 11, 21 | 55, | 1.046066265, | 21, 34 |
| 10, | 1.004899090, | 3, 7 | 33, | 1.024188296, | 12, 21 | 56, | 1.047164848, | 21, 35 |
| 11, | 1.005598870, | 3, 8 | 34, | 1.025087627, | 13, 21 | 57, | 1.048263370, | 21, 36 |
| 12, | 1.006298670, | 4, 8 | 35, | 1.026086758, | 13, 22 | 58, | 1.049361844, | 21, 37 |
| 13, | 1.006998450, | 5, 8 | 36, | 1.027085839, | 13, 23 | 59, | 1.050460296, | 21, 38 |
| 14, | 1.007798130, | 5, 9 | 37, | 1.028084881, | 13, 24 | 60, | 1.051558689, | 21, 39 |
| 15, | 1.008597780, | 5, 10 | 38, | 1.029083911, | 13, 25 | 61, | 1.052657102, | 21, 40 |
| 16, | 1.009397411, | 5, 11 | 39, | 1.030082893, | 13, 26 | 62, | 1.053755455, | 21, 41 |
| 17, | 1.010197051, | 5, 12 | 40, | 1.031081904, | 13, 27 | 63, | 1.054853758, | 21, 42 |
| 18, | 1.010996661, | 5, 13 | 41, | 1.032080865, | 13, 28 | 64, | 1.055952151, | 21, 43 |
| 19, | 1.011796321, | 6, 13 | 42, | 1.033079786, | 13, 29 | 65, | 1.057050485, | 21, 44 |
| 20, | 1.012595951, | 7, 13 | 43, | 1.034078807, | 13, 30 | 66, | 1.058148768, | 21, 45 |
| 21, | 1.013395561, | 8, 13 | 44, | 1.035077779, | 13, 31 | 67, | 1.059247031, | 21, 46 |
| 22, | 1.014295032, | 8, 14 | 45, | 1.036076710, | 13, 32 | 68, | 1.060345235, | 21, 47 |
| 23, | 1.015194462, | 8, 15 | 46, | 1.037075631, | 13, 33 | 69, | 1.061443638, | 21, 48 |
| 24, | 1.016093863, | 8, 16 | 47, | 1.038074503, | 13, 34 | 70, | 1.062541983, | 21, 49 |
| 25, | 1.016993263, | 8, 17 | 48, | 1.039073584, | 14, 34 | 71, | 1.063640278, | 21, 50 |
| 89, | 1.083409245, | 34, 55 | 144, | 1.14925821, | 55, 89 | 233, | 1.26456020, | 89, 144 |
| 377, | 1.46511596 | 144, 233 | 610, | 1.81188758, | 233, 377 | 987, | 2.40799356, | 377, 610 |
| 1597, | 3.426668, | 610, 987 | 2584, | 5.1563753, | 987, 1597 | 4181, | 8.072368, | 1597, 2584, |
| 6765, | 12.948090 | 2584, 4181 | | | | | | |

As stated above, the table assumes $q = .9999$ for various group sizes, $n$, whether that group is an entire population or situated within a larger structure. The second column gives the expected number of tests required to locate all positive samples in that group, using the optimal search algorithm. The third column gives the size of the two subgroups to be tested if the group tests positive. By following the division into successive subgroups one can obtain the full binary structure which represents the optimal search. For example, if during the course of a procedure a group of size 233 occurs and tests positive, the subsequent subgroups should be of sizes 89 and 144; if the subgroup of 89 tests positive, further subgroups of 34 and 55 are tested, and so on until subgroups testing negative are eliminated and only individual positive samples remain. When applied to a population of one million, the expected number of tests using this procedure is 1913.982. This compares favorably with the results of the above mentioned student procedures that assume there are exactly 100 positive samples, rather than the probability p = .0001. (In an actual test situation this would be adjusted slightly, since 6765 does not divide exactly into one million. We will omit this refinement.)

## The Fibonacci conjecture

From the output of hundreds of computer runs it appears that the optimal groupings are always the same, regardless of $q$. For instance, in an optimal structure, a test that includes 55 samples is always followed by tests that include 21 and 34 samples, and a test of 34 samples is followed by tests that include 13 and 21 samples. An inspection of Table 1 suggests a Fibonacci based pattern.

CONJECTURE: *If the optimal fixed structure calls for the inclusive test of a group of size n, then if n is the Fibonacci number $F_k$ , the two subgroups tested within this group will be of sizes $F_{k-1}$ and $F_{k-2}$ . If n is not a Fibonacci number, then these two subgroups will be of sizes m and (n − m), where at least one of these is a Fibonacci number, and there is exactly one Fibonacci number between m and (n–m); there will be only one pair that satisfies the conditions.*

Although the conjecture has been verified for 20 different values of $q$, it has not been proven.

We note that the running time to implement the Fibonacci conjecture is considerably shorter than the running time for the optimal search algorithm presented above.

## Adjacent remarks

**Concerning the Fibonacci conjecture**    Neither a proof of the Fibonacci conjecture, nor a counterexample, has as yet emerged. There is a superficial resemblance to two results concerning binary searches: the "golden section search procedure" in non-linear programming [**2**, pp. 179-183, **6**], and the study of Fibonacci trees in [**3**]. Yet they start from different assumptions and do not support any proof.

**A broader definition of a valid search**    Definition 1 assumes a valid search procedure which is completely predetermined except for the omission of a test whose result is already known. This concise definition might be relaxed to accommodate the following common occurrence. Assume that we have reached a stage at which we are testing $\overline{ab}$, where $a$ and $b$ are substructures. If together they test positive, and then $a$ alone tests positive, we now know nothing about $b$. Indeed, we know nothing about any sample or group to the right of $b$, so that instead of proceeding with our specified program—which requires us to test $b$—it would be more economical to begin again with all samples about which nothing is known. That is, we would apply the algorithm of Theorem 4 to all samples to the right of $a$, creating a new structure, the number of whose tests we would add to those we already obtained. This would be done at every further occurrence of an $\overline{ab}$ situation. Under this procedure of constant restructuring the expected number of tests for the above example would be reduced from 1913.982 to 1542.691. This result is comparable to those of other procedures applied to the more restrictive version of the problem that assumes exactly 100 positive samples [**4**].

It should be emphasized that in spite of such results, the algorithm of Theorem 4 has not been shown to be the optimal algorithm to apply in a constantly restructuring procedure.

**An optimal procedure without the assumption of nesting**    While every study, including this one, has assumed that testing proceeds in a nested manner, this does not necessarily produce the optimal result. For a population as small as $n = 3$, if $q > \dfrac{1 + \sqrt{33}}{8} \approx .843$, there is a non-nested procedure which gives a lower expected number of tests than the algorithm given above. However, the physical application of a non-nested algorithm might be impractical, as, for example, in searches within fixed circuitry.

## Proofs

Several definitions of convenience will be made within the proofs, often using the sign "$\equiv$".

**Proof of Theorem 2**   By Theorem 1, the only test affected by the existence or non-existence of the proposed test is the test immediately above it. Thus we need consider only a structure such as $\overline{\overline{n_j} \dots \overline{\overline{n_k} \dots \overline{n_l}} \dots \overline{n_m}}$, where, in keeping with Theorem 1, we've noted the number of samples in groups $j,k,l$ and $m$. We assume that the groups are actually numbered successively, so that $j < \dots < k < \dots < l < \dots < m$, and a difference such as $m - j$ equals the exact number of groups between $m$ and $j$ plus one. The proposed, immediately disadvantageous test, includes groups $k$ through $l$, and the test of all the groups from $j$ through $m$ is the next test outside it. There are two cases:

Case 1.   Group $l$ is not group $m$; that is, there are groups between them;   equivalently, $l - m > 0$. In this case the test $\overline{\overline{n_j} \dots \overline{n_m}}$ *without* $\overline{\overline{n_k} \dots \overline{n_l}}$ adds $1 - q^{\sum\limits_{j}^{m-1} n_i} - (m - j) q^{\sum\limits_{j}^{m} n_i} \equiv E_2$

$$(2)$$

where $\Sigma$ indicates summation through the groups from left to right.

The test $\overline{\overline{n_j} \dots \overline{n_m}}$ *with* $\overline{\overline{n_k} \dots \overline{n_l}}$ adds $1 - q^{\sum\limits_{j}^{m-1} n_i} - (m - j - (l - k)) q^{\sum\limits_{j}^{m} n_i} \equiv E_3.$ $\qquad(3)$

$E_2 - E_3 \;=\; -(l - k) q^{\sum\limits_{j}^{m} n_i} \equiv E_4$ and $E_4 < 0.$ $\qquad\qquad(4)$

Since $\overline{\overline{n_k} \dots \overline{n_l}}$ was assumed to be disadvantageous, the total addition without $\overline{\overline{n_k} \dots \overline{n_l}}$ is better (lower) than with it.

Case 2. Group $l$ is the same as group m; that is, $l - m = 0$. In this case $\overline{\overline{n_j ... n_m}}$ without $\overline{\overline{n_k ... n_l}}$ is the same as in *Case 1.* $\overline{\overline{n_j ... n_m}}$ with $\overline{\overline{n_k ... n_l}}$ adds

$$1 - q^{\sum\limits_{j}^{k-1} n_i} - (m - j - (m - k))q^{\sum\limits_{j}^{m} n_i} = 1 - q^{\sum\limits_{j}^{k-1} n_i} - (k - j)q^{\sum\limits_{j}^{m} n_i} \equiv E_5. \tag{5}$$

$$E_2 - E_5 = q^{\sum\limits_{j}^{k-1} n_i} + (k - j)q^{\sum\limits_{j}^{m} n_i} - q^{\sum\limits_{j}^{m-1} n_i} - (m - j)q^{\sum\limits_{j}^{m} n_i}$$

$$= q^{\sum\limits_{j}^{k-1} n_i} [1 - q^{\sum\limits_{k}^{m-1} n_i} - (m - k)q^{\sum\limits_{k}^{m} n_i}] \equiv E_6, \tag{6}$$

which, since $l = m$, is exactly a fraction ($< 1$) of the amount added by the disadvantageous test $\overline{\overline{n_k ... n_l}}$. But since the entire structure *with* $\overline{\overline{n_k ... n_l}}$ must also add on the positive value of $\overline{\overline{n_k ... n_l}}$, this is more than the fractional advantage it has over the structure *without* $\overline{\overline{n_k ... n_l}}$. ∎

## Two Lemmas for the proof of Theorem 3

LEMMA 1. *Given the structure* $\overline{\overline{n_1 n_2 n_3 .... n_{k-1} n_k}}$, *where* $n_k \geq n_1 \geq n_2 \geq n_3 \geq . . . n_{k-1}$ , *and where the inclusive test is advantageous, then at least one of the following is better than the given structure:*

$$A = \overline{\overline{\overline{n_1 n_2 n_3 .... n_{k-1} n_k}}} \quad or \quad B = \overline{\overline{\overline{n_1 n_2 n_3 .... n_{k-1} n_k}}} .$$

Proof: Above the k groups in the given structure, the addition of the overall (advantageous) test adds on the negative value $1 - q^{n - n_k} - (k - 1)q^n \equiv E_7.$ \tag{7}

For structure $B$, the addition of the inclusive test adds:

$$[1 - q^{n - n_1 - n_k} - (k - 2)q^{n - n_1}] + [1 - q^{n_1} - q^n] \equiv E_8. \tag{8}$$

For structure $A$, the addition of the inclusive test adds:

$$[1 - q^{n_{k-1}} - q^{n_{k-1} + n_k}] + [1 - q^{n - n_{k-1} - n_k} - (k - 2)q^n] \equiv E_9. \tag{9}$$

Thus we have to prove that either $E_8 - E_7 < 0$ or $E_9 - E_7 < 0$.

$$E_8 - E_7 = 1 - q^{n - n_1 - n_k} - (k - 2)q^{n - n_1} + 1 - q^{n_1} - q^n - 1 + q^{n - n_k} + (k - 1)q^n$$

$$= 1 - q^{n - n_1 - n_k} - (k - 2)q^{n - n_1} - q^{n_1} + q^{n - n_k} + (k - 2)q^n$$

$$= (1 - q^{n_1})(1 - q^{n - n_k - n_1} - (k-2)q^{n - n_1}) \quad \equiv E_{10}, \tag{10}$$

which is $(1 - q^{n_1})$ times the amount added by $\overline{\overline{n_2 \dots n_k}}$.

Define $(1 - q^{n - n_k - n_1} - (k-2)q^{n - n_1})$, the amount added by $\overline{\overline{n_2 \dots n_k}}$, as $E_{11}$. (11)

If $E_{11}$ is negative, then $E_8$ is better than $E_7$ and the proof is finished. Assume, therefore, that $E_{11}$ is positive, and consider $E_9 - E_7$ under this assumption.

$$
\begin{aligned}
E_9 - E_7 &= 1 - q^{n_{k-1}} - q^{n_{k-1} + n_k} + 1 - q^{n - n_{k-1} - n_k} - (k-2)q^n - 1 + q^{n - n_k} + (k-1)q^n \\
&= 1 - q^{n_{k-1}} - q^{n_{k-1} + n_k} - q^{n - n_{k-1} - n_k} + q^{n - n_k} + q^n \\
&= 1 - q^{n_{k-1}} - q^{n_{k-1} + n_k} - q^{n - n_{k-1} - n_k}(1 - q^{n_{k-1}} - q^{n_{k-1} + n_k}) \\
&= (1 - q^{n - n_{k-1} - n_k})(1 - q^{n_{k-1}} - q^{n_{k-1} + n_k}) \equiv E_{12}, \tag{12}
\end{aligned}
$$

which is $(1 - q^{n - n_{k-1} - n_k})$ times the amount added by $\overline{\overline{n_{k-1} n_k}}$.

Let $\quad (1 - q^{n_{k-1}} - q^{n_{k-1} + n_k}) \equiv E_{13}$. (13)

If $E_{13}$ is negative, then $E_9$ is better than $E_7$. Assume, therefore, that $E_{13}$ is positive.

We will now show that the assumption that both $E_{11}$ and $E_{13}$ are positive leads to a contradiction, which will prove Lemma 1. Since $E_7$ has been assumed negative, to show that $E_{11}$ is in fact negative, contrary to our assumption, we need only show that $E_7 - E_{11}$ is positive:

$$
\begin{aligned}
E_7 - E_{11} &= 1 - q^{n - n_k} - (k-1)q^n - 1 + q^{n - n_k - n_1} + (k-2)q^{n - n_1} \\
&= q^{n - n_k - n_1} + (k-2)q^{n - n_1} - q^{n - n_k} - (k-1)q^n \\
&= q^{n - n_k - n_1}[1 + (k-2)q^{n_k} - q^{n_1} - (k-1)q^{n_1 + n_k}] \\
&= q^{n - n_k - n_1}[1 - q^{n_1} - q^{n_k + n_1}] + q^{n - n_k - n_1}[(k-2)q^{n_k} - (k-2)q^{n_1 + n_k}] \equiv E_{14}. \tag{14}
\end{aligned}
$$

Now, we have assumed that $n_1 \geq n_{k-1}$ and therefore the first quantity in brackets in (14) is at least as great as $E_{13}$. Since $E_{13}$ was assumed positive, and since the second bracketed quantity in (14) is positive, $E_{14}$ must be positive. The contradiction proves Lemma 1. ∎

**LEMMA 2** *In Lemma 1, the structure* $A = \overline{\overline{n_1 n_2 n_3}} \dots \overline{\overline{n_{k-1} n_k}}$ *is always advantageous over the original structure* $\overline{\overline{n_1 n_2 n_3}} \dots \overline{\overline{n_{k-1} n_k}}$.

**Proof:** Since Lemma 1 proved that at least one of $A$ or $B$ is always advantageous over the original $\overline{\overline{n_1 n_2 n_3}} \dots \overline{\overline{n_{k-1} n_k}}$, we need only show that $B$ advantageous implies $A$ advantageous. This is true if $E_{11}$ negative implies $E_{12}$ negative. Assuming that $E_{13}$ is positive, and remembering that $n_k \geq n_1 \geq \ldots \geq n_{k-1}$, we now compute the value of $\overline{n_1 n_2} \ldots \overline{\overline{n_{j-1} \ldots n_k}}$ *minus* the value of $\overline{n_1 n_2} \ldots \overline{\overline{n_j \ldots n_k}}$. This difference equals:

$$[1 - q^{\sum\limits_{j-1}^{k-1} n_i} - (k-j+1)q^{\sum\limits_{j-1}^{k} n_i}] - [1 - q^{\sum\limits_{j}^{k-1} n_i} - (k-j)q^{\sum\limits_{j}^{k} n_i}]$$

$$= -q^{\sum\limits_{j-1}^{k-1} n_i} - (k-j+1)q^{\sum\limits_{j-1}^{k} n_i} + q^{\sum\limits_{j}^{k-1} n_i} + (k-j)q^{\sum\limits_{j}^{k} n_i}$$

$$= q^{\sum\limits_{j}^{k-1} n_i}[-q^{n_{j-1}} - (k-j+1)q^{n_k + n_{j-1}} + 1 + (k-j)q^{n_k}]$$

$$= q^{\sum\limits_{j}^{k-1} n_i}[1 - q^{n_{j-1}} - (k-j+1)q^{n_k + n_{j-1}} + (k-j)q^{n_k}] \equiv E_{15}. \qquad (15)$$

The factor in brackets in $E_{15}$ is $\geq [1 - q^{n_{j-1}} - (k-j+1)q^{n_k + n_{j-1}} + (k-j)q^{n_k + n_{j-1}}]$, which simplifies to $[1 - q^{n_{j-1}} - q^{n_k + n_{j-1}}]$, which itself is $\geq [1 - q^{n_{k-1}} - q^{n_k + n_{k-1}}]$ since $n_{j-1} \geq n_{k-1}$. This last term, $[1 - q^{n_{k-1}} - q^{n_k + n_{k-1}}]$, is exactly $E_{13}$, however, and therefore we know that as the test is extended one more group to the left it grows more positive. Since $E_{11}$ is eventually reached by such shifts to the left, it will be positive. Therefore a negative $E_{11}$ implies a negative $E_{12}$, and Lemma 2 is proved. ∎

**Rewriting and proof of Theorem 3**  Three definitions are needed:

DEFINITION 8. *If a test has n − 1 tests that include it, it is said to be on level n.*

DEFINITION 9. *If a test includes s groups immediately beneath it, it is said to have an excess of s − 2.*

DEFINITION 10. *A test is said to have an excess if its excess is ≥ 1.*

We now rewrite Theorem 3 in a form that we can prove more conveniently:

THEOREM 3, rewritten.  *If a structure contains any test with an excess, there is another structure on the same samples, none of whose tests has an excess, and which is at least as good as the given structure.*

Proof:   The following is a method for changing the original structure into one without excess. At any stage take a test $A$ with excess, on the lowest possible level. Arrange the groups so that $n_k \geq n_1 \geq n_2 \geq n_3 \geq \ldots n_{k-1}$. Since the value of $A$ is $1 - q^{n-n_k} - (k-1)q^n$, by placing the largest group (maximum $n_i$) to the right we can only be working to our advantage. Put a test over $\overline{n_{k-1}n_k}$ so that $\overline{\overline{n_{k-1}n_k}}$ is now part of the structure. By Lemma 2 this is advantageous. If $A$ is still negative, continue. The excess within $A$ will either be eaten away in this manner, or, at some point, $A$ will turn positive. If this happens, by Theorem 2 we can erase the test $A$ without disadvantage, and the remaining excess is lifted to a higher level. In fact, if the next test above $A$ is made positive by the removal of $A$, by Theorem 2 we can drop that test also, and the excess will go to an even higher level. In this way, all the excesses are eliminated.                                                             ∎

**Proof of Theorem 4, the optimal algorithm**   We see immediately why we need only go up to $[n/2]$. For the possibilities in $A$ are independent of the order of $n'$ and $n-n'$, and those in $B$ are more advantageous if the smaller group is placed to the left. That the procedure is optimal we see as follows: Consider $O_q(n)$, the optimal structure on $n$ samples. If it has no inclusive test, then it is the sum of two smaller structures (one or both of which might again have no inclusive test.) The minimum of the possibilities of type $A$ will clearly give us this. If $O_q(n)$ has an inclusive test, by Theorem 3 the test cannot include more than two substructures within it, so that all permissible cases are included in $B$. That we always want to use $O_q(n')$ and $O_q(n-n')$ can be seen from the fact that since the test over all n samples is independent of what exists within the next layer inside, we would certainly never want to substitute for a fully tested $O_q(n')$ or $O_q(n-n')$ another fully tested structure on $n'$ samples or $(n-n')$ samples. Furthermore, by Theorem 2, if $O_q(n')$ or $O_q(n-n')$ were not fully tested, we would never want to add a disadvantageous test for the purpose of gaining an advantage with a test of all $n$ samples.                                                             ∎

**Summary**  The use of group testing to locate all instances of disease in a large population of blood samples was first considered more than seventy years ago.  Since then several procedures have been used to lower the expected number of tests required. The algorithm presented here, in contrast to previous ones, takes a constructive rather than a top-down approach. As far as could be verified, it offers the first proven solution to the problem of finding the minimum expected number of tests using a predetermined procedure. Computer results strongly suggest that the algorithm has a Fibonacci-based pattern.